\documentclass[fleqn,usenatbib]{mnras}

\usepackage{xspace}
\usepackage{mathptmx}

\usepackage[T1]{fontenc}

\usepackage{lineno}

\DeclareRobustCommand{\VAN}[3]{#2}
\let\VANthebibliography\thebibliography
\def\thebibliography{\DeclareRobustCommand{\VAN}[3]{##3}\VANthebibliography}

\usepackage{graphicx}	
\usepackage{amsmath}	
\usepackage{xcolor}
\usepackage{algorithm}
\usepackage{algpseudocode}
\usepackage{enumitem}

\def\lsi{LS\,I\,$+$61\,303\xspace}
\def\pulx{Swift\,J0243.6$+$6124\xspace}
\newcommand{\degrees}{\ensuremath{^\circ}}
\newcommand{\dotdegrees}{\ensuremath{\overset{\degrees}{.}}}

\title[Baikal-GVD neutrinos and high-energy sources]{Search for directional associations between Baikal Gigaton Volume Detector neutrino-induced cascades and high-energy astrophysical sources}

\author[]{\parbox{\textwidth}{
V.A.~Allakhverdyan,$^{1}$
A.D.~Avrorin,$^{2}$
A.V.~Avrorin,$^{2}$
V.~M.~ Aynutdinov,$^{2}$
Z.~Barda\v{c}ov\'{a},$^{3,6}$
I.A.~Belolaptikov,$^{1}$
E.A.~Bondarev,$^{2}$
I.V.~Borina,$^{1}$
N.M.~Budnev,$^{4}$
A.S.~Chepurnov,$^{8}$
V.Y.~Dik,$^{1,5}$
G.V.~Domogatsky,$^{2}$
A.A.~Doroshenko,$^{2}$
R.~Dvornick\'{y},$^{3}$
A.N.~Dyachok,$^{4}$
Zh.-A.M.~Dzhilkibaev,$^{2}$
E.~Eckerov\'{a},$^{3,6}$
T.V.~Elzhov,$^{1}$
L.~Fajt,$^{6}$
A.R.~Gafarov,$^{4}$
K.V.~Golubkov,$^{2}$
N.S.~Gorshkov,$^{1}$
T.I.~Gress,$^{4}$
K.G.~Kebkal,$^{7}$
I.~Kharuk,$^{2}$
E.V.~Khramov,$^{1}$
M.M.~Kolbin,$^{1}$
K.V.~Konischev,$^{1}$
A.V.~Korobchenko,$^{1}$
A.P.~Koshechkin,$^{2}$
V.A.~Kozhin,$^{8}$
M.V.~Kruglov,$^{1}$
V.F.~Kulepov,$^{9}$
Y.E.~Lemeshev,$^{4}$
M.B.~Milenin,$^{2}$\thanks{Deceased}
R.R.~Mirgazov,$^{4}$
D.V.~Naumov,$^{1}$
A.S.~Nikolaev,$^{8}$
D.P.~Petukhov,$^{2}$
E.N.~Pliskovsky,$^{1}$
M.I.~Rozanov,$^{10}$
E.V.~Ryabov,$^{4}$
G.B.~Safronov,$^{2}$
D.~Seitova,$^{1,5}$
B.A.~Shaybonov,$^{1}$
M.D.~Shelepov,$^{2}$
S.D.~Shilkin,$^{2}$
E.V.~Shirokov,$^{8}$
F.~\v{S}imkovic,$^{3,6}$
A.E. Sirenko,$^{1}$
A.V.~Skurikhin,$^{8}$
A.G.~Solovjev,$^{1}$
M.N.~Sorokovikov,$^{1}$
I.~\v{S}tekl,$^{6}$
A.P.~Stromakov,$^{2}$
O.V.~Suvorova,$^{2}$
V.A.~Tabolenko,$^{4}$
B.B.~Ulzutuev,$^{1}$
Y.V.~Yablokova,$^{1}$
D.N.~Zaborov$^{2}$
S.I.~Zavyalov,$^{1}$
D.Y.~Zvezdov,$^{1}$
(Baikal-GVD Collaboration), \\
N.A.~Kosogorov,$^{11,12,17}$\thanks{Corresponding author, e-mail: nakosogorov@gmail.com}
Y.Y.~Kovalev,$^{13,11,12}$
G.V.~Lipunova,$^{14,13}$
A.V.~Plavin,$^{12}$
D.V.~Semikoz,$^{15}$
S.V.~Troitsky$^{2,16}$}
\vspace{0.25cm}\\
\parbox{\textwidth}{
$^{1}$Joint Institute for Nuclear Research, Dubna, 141980  Russia\\
$^{2}$Institute for Nuclear Research of the Russian Academy of Sciences, 60th October Anniversary Prospect 7a, Moscow 117312, Russia\\
$^{3}$Comenius University, Bratislava, 81499 Slovakia\\ 
$^{4}$Irkutsk State University, Irkutsk, 664003 Russia\\ 
$^{5}$Institute of Nuclear Physics ME RK, Almaty, 050032 Kazakhstan\\ 
$^{6}$Czech Technical University in Prague, Institute of Experimental and Applied Physics, 11000 Prague, Czech Republic\\ 
$^{7}$AO "LATENA" (Joint Stock Company), St. Petersburg, 199106 Russia\\ 
$^{8}$Skobeltsyn Research Institute of Nuclear Physics, Lomonosov Moscow State University, Moscow, 119991 Russia\\ 
$^{9}$Nizhny Novgorod State Technical University, Nizhny Novgorod, 603950 Russia\\ 
$^{10}$St. Petersburg State Marine Technical University, St. Petersburg, 190008 Russia\\ 
$^{11}$Astro Space Center of Lebedev Physical Institute, Profsoyuznaya 84/32, 117997 Moscow, Russia\\ 
$^{12}$Moscow Institute of Physics and Technology, Institutsky per. 9, Dolgoprudny 141700, Russia\\ 
$^{13}$Max-Planck-Institut f\"ur Radioastronomie, Auf dem H\"ugel 69, 53121 Bonn, Germany\\ 
$^{14}$Sternberg Astronomical Institute,  Lomonosov Moscow State University, Universitetskii pr. 13, Moscow, 119234 Russia\\
$^{15}$APC, Universit\'e Paris Diderot, CNRS/IN2P3, CEA/IRFU, Sorbonne Paris Cit\'e, 119 75205 Paris, France\\ 
$^{16}$Physics Department,  Lomonosov Moscow State University, 1-2 Leninskie Gory,  Moscow 119991, Russia\\
$^{17}$Cahill Center for Astronomy and Astrophysics, MC 249-17 California Institute of Technology, Pasadena, CA 91125, USA}}

\date{Accepted 2023 August 24. Received 2023 August 22; in original form 2023 July 12}

\pubyear{2023}

\begin{document}
\label{firstpage}
\pagerange{\pageref{firstpage}--\pageref{lastpage}}
\maketitle

\begin{abstract}
\normalsize
{Baikal-GVD has recently published its first measurement of the diffuse astrophysical neutrino flux, performed using high-energy cascade-like
events. We further explore the Baikal-GVD cascade dataset collected in
2018-2022, with the aim to identify possible associations between the
Baikal-GVD neutrinos and known astrophysical sources.
We leverage the relatively high angular resolution of the Baikal-GVD neutrino telescope (2-3 deg.), made possible by the use of liquid water as the detection medium, enabling the study of astrophysical point sources even with cascade events. We estimate the telescope's sensitivity in the cascade channel for high-energy astrophysical sources and refine our analysis prescriptions using Monte-Carlo simulations. We primarily focus on cascades with energies exceeding 100 TeV, which we employ to search for correlation with radio-bright blazars. Although the currently limited neutrino sample size provides no statistically significant effects, our analysis suggests a number of possible associations
with both extragalactic and Galactic sources. Specifically, we present an analysis of an observed triplet of neutrino candidate events in the Galactic plane, focusing on its potential connection with certain Galactic sources, and discuss the coincidence of cascades with several bright and flaring blazars.
}

\end{abstract}




\section{Introduction}

In recent years, high-energy neutrinos have become a hot topic of astrophysical research. Various sources and processes have been discussed in which these neutrinos could be produced efficiently. The IceCube detector \citep{2013Sci...342E...1I} demonstrated that the arrival directions of these high-energy neutrinos are distributed mostly isotropically over the sky \citep{2014PhRvL.113j1101A}. While subsequent studies found evidence for a certain Galactic component \citep{neutgalaxy,ANTARES-ridge,IC_Galaxy2023}, most of the neutrinos are still expected to arrive from extragalactic sources, like active galactic nuclei (AGN), gamma-ray bursts, tidal disruption events, etc.\ \citep[e.g.][]{2017ARNPS..67...45M, 2021PhyU...64.1261T}. Based on recent studies, AGNs turned out to be very promising sources of neutrino production. For example, the directional coincidence between the neutrino event 170922A detected by IceCube and the blazar TXS~0506$+$056 as well as its temporal coincidence with a $\gamma$-ray flare from the source demonstrated a significant blazar-neutrino association \citep{2018Sci...361.1378I,2018Sci...361..147I}. Recently a nearby active galaxy NGC~1068 was also singled out as a probable neutrino source \citep{doi:10.1126/science.abg3395}. 

Statistical studies of very-long-baseline interferometry (VLBI) sources showed the presence of significant correlation between bright blazars and IceCube neutrino events \citep{2020ApJ...894..101P,2021ApJ...908..157P,Plavin2023}. 
It should be noted that the conclusions from multiple analyses done by different authors deliver varied outcomes, some confirming and some questioning or refuting the presence of the correlation. Typically, these conclusions depend on the sample of blazars and type of neutrino data under investigation. Compare \citet{2021PhRvD.103l3018Z,hovatta2021,2022ApJ...933L..43B,2023arXiv230511263B,2023arXiv230412675A}. 
Clearly, more observing data and even stronger statistics are needed.
Relativistic jets seem to be a great candidate for neutrino emission due to their effective particle acceleration. Such acceleration can produce high-energy protons, and as a consequence generate neutrinos \citep[e.g.][]{2018ApJ...865..124M,MuraseTXS,kun2021,2021MNRAS.503.3145B,2023MNRAS.519.1396S}. 
On the other hand, neutrinos can be produced in various Galactic sources of high-energy radiation, see e.g.\ \citet{Kheirandish,2021PhyU...64.1261T}.

With a mission to establish the sources of neutrinos, the Baikal Gigaton Volume Detector \citep[Baikal-GVD; see e.g.][]{2018arXiv180810353B, 2019EPJWC.20701003A} operates as a large-scale Cherenkov detector tracking neutrinos in the TeV-PeV energy band.
In this work we search for directional coincidence between such neutrinos and extragalactic as well as Galactic sources. The first analysis of data from four years of high-energy cascade-like event observations, conducted by the Baikal-GVD neutrino telescope, was presented in \citet{Baikal-diffuse}. The analysis confirmed the IceCube telescope's discovery of a diffuse astrophysical neutrino flux \citep{2013Sci...342E...1I}. The new astrophysical neutrino candidates, selected using Baikal-GVD, are extremely intriguing from the perspective of identifying their sources.
These early Baikal-GVD cascade data are further analyzed in this work.
In particular, a statistical correlation between the neutrino candidate
events and a VLBI-bright blazar sample is searched for, similarly to the
analysis in \citet{2020ApJ...894..101P,Plavin2023}.
Apart from that, we discuss how many events will be needed to reach a certain significance level. The analysis procedure was structured similar to a blind analysis, in the sense that the analysis was not tailored to fit the actual data. We then examine the actual statistics and carry out an analysis of radio flux density correlation. Lastly, we briefly describe the most interesting cascades in our sample and their potential sources.

We describe the data in \autoref{data}. In \autoref{analysis}, we study correlations between the Baikal-GVD dataset and a sample of blazars and discuss the most notable coincident Galactic and extragalactic sources. Specifically, we describe all the algorithms in \autoref{algorithms}, investigate a possibility of finding any correlations in \autoref{blind} and examine the actual statistics based on the real data in \autoref{unblinding}. We then describe the most interesting potential neutrino emitters in \autoref{special}. Finally, \autoref{sum} sums up all the results.

\section{Data}
\label{data}
\subsection{Baikal-GVD cascades}
\label{neu_data}

The Baikal-GVD is a cubic-kilometer scale neutrino observatory located in Lake Baikal ($51^\circ50^\prime$ N, $104^\circ20^\prime$ E). The lake depth at the facility site is 1366 m. Basic elements of Baikal-GVD are photomultiplier tubes (PMTs) and related electronics contained within pressure-resistant glass spheres; they are referred to as optical modules (OMs). OMs are mounted onto vertical cables forming “strings”. Each string comprises 36 OMs spaced $\sim15$ m apart vertically at depths from 750~m to 1275~m. Strings with OMs are collected in clusters. Each cluster is an independent array comprising 8 strings with a total of 288 OMs and is connected to the shore station by its own electro-optical cable. Seven of these eight strings are arranged in a heptagonal grid with $\sim60$~m spacing. Inter-cluster distances between the central strings vary from 250 to 300~m. The instrument has been operating since 2016 with the effective volume increasing every year. The current rate of array deployment is about two clusters per year. The operational configuration of Baikal-GVD in 2022 consisted of 10 clusters and comprised a total of 2916 OMs. 

Baikal-GVD detects Cherenkov radiation of charged particles, which are produced in neutrino interactions. The Cherenkov radiation detected by the light sensors forms two types of patterns, tracks and cascades. In the case when the incident particle is a muon neutrino, a muon is produced in the charged-current (CC) interaction, which generally traverses several kilometers in water or bed-rock and exit the kilometer-scale detection volume producing a track-like event. Neutral-current (NC) interactions of all types of neutrinos, as well as most of CC interactions of electron and tau neutrinos, yield hadronic and electromagnetic showers (cascades). The showers are quasi point-like, highly anisotropic sources of Cherenkov radiation. The energy of the cascade progenitor neutrino is determined with a good accuracy; however, the angular reconstruction is worse for cascades than for tracks. The cascade channel is thus complementary to the track one.

In this work, we study Baikal-GVD cascade-like neutrino candidate events. The events of this type are observed as showers of charged particles, with the accuracy of cascade energy reconstruction being relatively high (about 10\%--30\%). At the same time, the direction accuracy is moderate and equals to about 2--4 degrees depending on the location and orientation of a cascade~\citep[see more detail in][]{2022JETP..134..399A}. 

We use Baikal-GVD data collected between April 2018 and March 2022. The telescope was operated in the configuration with 3 clusters in 2018--2019, 5 clusters in 2019--2020, 7 clusters in 2020--2021, while from April 2021 to March 2022 the telescope comprised 8 clusters.

We have performed data analysis for individual clusters as independent setups. The data
sample comprises $3.49\times10^{10}$ events, collected using the basic trigger criteria of the telescope. 
After applying noise hit suppression, cascade reconstruction and cuts on
reconstruction quality parameter, as described in detail in \citet{2019EPJWC.20705001A} and \citet{Baikal-diffuse}, also with condition on
OM hit multiplicity $N_{\rm hit}$ > 12, the sample of 6157 cascades with
reconstructed energy ranged from 10 TeV to 1200 TeV was selected. The upper limit corresponds to the highest energy cascade in our sample.

Two subsamples of neutrino events were selected from the cascade sample \citep[][]{Baikal-diffuse}. One of them comprises high-energy cascades from all sky directions with OM hit multiplicity $N_{\rm hit}$ > 19 and reconstructed energy $E_{\rm sh} \geq 100$ TeV. Parameters of these 13 events are listed in \autoref{tab:HE_events}. The other sample comprises upward moving cascades with OM hit multiplicity $N_{\rm hit} > 12$, reconstructed energy $E_{\rm sh} > 10$ TeV and
 zenith angles $\cos \theta < -0.25$. We also denote such events as under horizon cascades. Parameters of these 11 events are listed in \autoref{tab:hor_events}. These two samples were used by \citet{Baikal-diffuse} to determine the astrophysical neutrino flux from the Baikal-GVD data, thus confirming the IceCube discovery of extraterrestrial neutrinos with an independent experiment. We note that the tables presented herein are the same as the ones from the referenced article, however, the first one omits three events due to the added constraint of OM hit multiplicity $N_{\rm hit}$ > 19.

\begin{figure*}
\centering
\includegraphics[width=1\textwidth,trim=100 300 100 350]{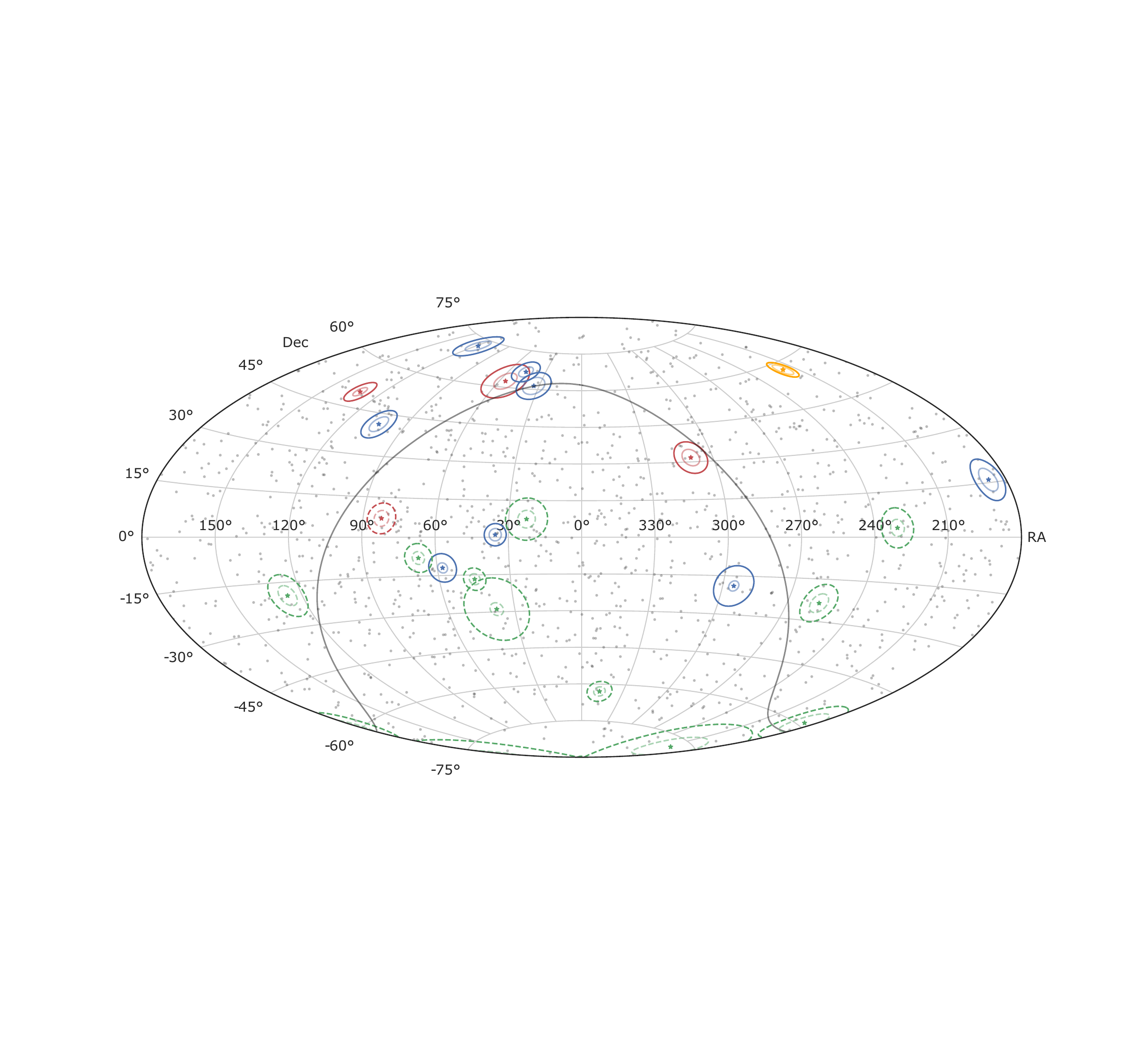}
\caption{Arrival directions of high-energy and under-horizon Baikal-GVD neutrino cascades (stars) and their uncertainty (50\% and 90\%) regions (ellipses) in the sky map (equatorial coordinates). Colour represents energies of the events: green is below 100 TeV, blue is between 100 TeV and 200 TeV, red is between 200 TeV and 1000 TeV, orange is above 1 PeV. Dashed ellipses show under-horizon events, while solid ellipses present events above the horizon. See \autoref{tab:HE_events} and \autoref{tab:hor_events} for more details. Positions of blazars from the 8 GHz VLBI sample with flux densities above 0.33 Jy are shown as grey dots. The Galactic plane is indicated as the grey curve. 
} 
\label{fig:HE_neu}
\end{figure*}

A skymap showing the high energy and under-horizon cascades with direction uncertainties is shown in \autoref{fig:HE_neu}, along with the positions of the VLBI blazars (see \autoref{data_vlbi}). The average direction
uncertainty for cascades with energies between 10 TeV and 100 TeV is
reported as a function of hit multiplicity in Table 3.

\begin{table*}
\centering
\caption{Parameters of the high-energy cascades: event ID, reconstructed energy, zenith angle, Galactic longitude and latitude, right ascension and declination, radii of the 50$\%$ and 90$\%$ uncertainty regions and alternative IDs from \citet{Belolaptikov:2021a1}. 
}
\begin{tabular}{|c|c|c|r|r|r|r|c|c|c|}
\hline
 Event ID & E & $\theta$ & $l$ & $b$ & RA & Dec & 50$\%$ unc. & 90$\%$ unc. & Alternative ID \\
 & TeV & deg. & deg. & deg. & deg. & deg. & deg. & deg. &  \\
\hline
   GVD181010CA  & 105 & 37   &  142.6   &  30.4   & 118.2 & 72.5    & 2.3 & 4.5 & GVD2018\_3\_354\_N\\
   GVD181024CA  & 115 & 73   &  164.1   &  $-$54.4   & 35.4  & 1.1     & 2.5 & 4.5 & GVD2018\_3\_383\_N \\
   GVD190216CA  & 398 & 64   &  141.4   &  5.8  & 55.6  & 62.4    & 3.3 & 6.9 & GVD2018\_2\_656\_N  \\
   GVD190517CA  & 1200& 61   &  99.9   &  54.9   & 217.7 & 57.6    & 2.0 & 3.0 & GVD2019\_2\_112\_N \\
   GVD190604CA  & 129 & 50   &  132.7   &  0.1   & 33.7  & 61.4    & 3.5 & 5.5 & GVD2019\_2\_153\_N \\
   GVD200826CA  & 110 & 71   &  21.0   &  $-$19.2   & 295.3 & $-$18.9 & 2.0 & 7.9 & GVD2020\_3\_175\_N \\
   GVD210117CA  & 246 & 57   &  168.8  &  38.8   & 131.9 & 50.2    & 1.6 & 3.6 & GVD2020\_6\_399\_N \\
   GVD210409CA  & 263 & 60   &  73.3   &  $-$6.1   & 310   & 31.7    & 3.3 & 6.3 & GVD2021\_7\_008\_N \\
   GVD210418CA & 224 & 115.5&  196.8     &  $-$14.6 & 82.4  & 7.1     & 3.0 & 5.8 & GVD2021\_1\_020\_N\\
   GVD210515CA  & 120 & 80.2 &  175.2  &  17.9    & 103.4 & 41.2    & 2.8 & 5.2 & GVD2021\_2\_048\_N \\
   GVD210716CA  & 110 & 58.7 &  135.5  &  7.1 & 46.0  & 66.7    & 2.1 & 4.1 & GVD2021\_4\_131\_N \\
   GVD210906CA  & 138 & 67.7 &  202.2 &  $-$45.3 & 57.8  & $-$12.0 & 2.0 & 5.6 & GVD2021\_6\_210\_N\\
   GVD220221CA  & 120 & 67.7 &  276.9  &  77.5    & 187.2 & 15.8    & 3.2 & 5.8 &  GVD2021\_1\_399\_N\\
\hline
    
\end{tabular}

\label{tab:HE_events}
\end{table*}

\begin{table*}
\centering
\caption{Parameters of the under-horizon cascade events with $\cos \theta < -0.25$. The columns are the same as in \autoref{tab:HE_events}. 
}

\begin{tabular}{|c|c|c|r|r|r|r|c|c|c|}
\hline
 Event ID & E & $\theta$ & $l$ & $b$ & RA & Dec & 50$\%$ unc. & 90$\%$ unc. & Alternative ID  \\
 & TeV & deg. & deg. & deg. & deg. & deg. & deg. & deg. &  \\
\hline
   GVD180504CA  & 25.1 & 111.7  &  299.1    &  3.6    & 185.4 & $-$59.0  & 3.9 & 6.9 & GVD2018\_1\_041\_N \\
   GVD190523CA  & 91.0 & 109.0  &  200.4   &  $-$58.4& 45.1  & $-$16.7  & 2.2 & 4.5 & GVD2019\_1\_114\_N  \\
   GVD200614CA  & 39.8 & 144.1  &  359.3     &  10.6   & 256.2 & $-$23.6  & 3.4 & 6.8 & GVD2020\_7\_079\_N  \\
   GVD201112CA  & 24.5 & 136.1  &  305.0    &  $-$15.1& 202.2 & $-$77.8  & 5.4 & 11.8& GVD2020\_3\_278\_N \\
   GVD210418CA  & 224 & 115.5&  196.8 &  $-$14.6 & 82.4  & 7.1     & 3.0 & 5.8 & GVD2021\_1\_020\_N \\
   GVD210501CA  & 63.1 & 112.3  &  223.4   &  $-$67.7& 38.1  & $-$28.9  & 2.6 & 12.6& GVD2021\_2\_030\_N \\
   GVD210506CA  & 21.9 & 114.2  &  5.9  &  46.7   & 230.6 & 3.1      & 2.8 & 6.6 & GVD2021\_5\_032\_N \\
   GVD210710CA  & 24.5 & 115.5  & 139.8&  $-$54.2& 22.7  & 7.4      & 3.6 & 8.6 & GVD2021\_7\_135\_N \\
   GVD210803CA  & 20.9 & 136.9  &  321.0    &  $-$50.3& 347.0 & $-$63.0  & 1.9 & 4.1 &  GVD2021\_6\_169\_N \\
   GVD220121CA  & 30.9 & 110.5  &  241.3   &  10.4   & 126.2 & $-$19.5  & 3.4 & 7.1 & GVD2021\_4\_355\_N  \\
   GVD220308CA  & 36.3 & 105.0  &  203.2   &  $-$35.2& 67.3  & $-$8.0   & 2.5 & 5.6 & GVD2021\_1\_409\_N  \\

\hline
    
\end{tabular}

\label{tab:hor_events}
\end{table*}

\begin{table}
	\centering
	\caption{The uncertainty of the reconstruction of the arrival directions of Baikal-GVD cascade events with energies between 10 and 100 TeV, averaged for all directions (2-nd column) and for ``under-horizon'' events (3-rd column). The numbers give the radii of 50\%-containment circles, in degrees, for different intervals of hit multiplicity.}
	\begin{tabular}{|l|c|c|}
		\hline
		$N_{\rm hit}$ & all & under horizon  \\ 
	
		\hline
        13--15 & 6.1 & 5.7 \\
        16--18 & 4.8 & 4.5 \\
        19--21 & 3.9 & 3.5 \\
        22--24 & 3.7 & 3.3 \\
        25--27 & 3.5 & 3.2 \\
        28--30 & 3.5 & 3.2 \\
        34--36 & 3.5 & 3.1 \\
        40--42 & 3.3 & 3.1 \\
        45--47 & 2.9 & 2.8 \\
        $>$50   & 2.8 & 2.6 \\
        \hline
	\end{tabular}

\label{tab:error}
\end{table}

\subsection{VLBI-selected sample of blazars}\label{data_vlbi}

We follow \citet{2020ApJ...894..101P,Plavin2023} and select 8 GHz VLBI observations of blazars,  which can be found in the Astrogeo\footnote{\url{http://astrogeo.org/vlbi_images/}}
database. These observations include geodetic VLBI programs \citep{2009JGeod..83..859P,2012A&A...544A..34P,2012ApJ...758...84P}, the Very Long Baseline Array (VLBA) calibrator
surveys \citep{2002ApJS..141...13B,2003AJ....126.2562F,2005AJ....129.1163P,2006AJ....131.1872P,2008AJ....136..580P,2007AJ....133.1236K,petrov2017,2016AJ....151..154G}, and other 8 GHz global VLBI, VLBA, EVN
(European VLBI Network), LBA (Australian Long Baseline Array) observations \citep{2011AJ....142...35P,2011AJ....142..105P,2011MNRAS.414.2528P,2012MNRAS.419.1097P,2013AJ....146....5P,2015ApJS..217....4S,2017ApJS..230...13S,2019MNRAS.485...88P}. These observations determine the positions and flux densities for blazars, which are publicly available from the Radio Fundamental Catalog (RFC)\footnote{\url{http://astrogeo.org/sol/rfc/rfc_2023a/}}.

We use the `blazar' term in this paper for active galaxies with Doppler-boosted jets pointing at an observer. Most of extragalactic radio sources with flat spectra are compact at parsec scales, while only about 25\,\% of steep spectrum sources show VLBI-detectable component in their emission \citep{2021AJ....161...88P}.
During the last dozens of years, VLBI surveys have observed compact extragalactic sources with both flat and steep radio spectrum. All of them are collected in the RFC.
We use the complete flux-density limited sample of extragalactic radio sources with integrated flux density at VLBI scales at 8~GHz from the RFC greater than 150~mJy. It is preliminary estimated to be 90-95\,\% complete on the basis of its contraction. 
More details on the RFC sample properties will be presented by Petrov \& Kovalev (in prep.).
Observations of VLBI-selected samples of extragalactic radio sources have shown that about 90\,\% of them are highly Doppler boosted jets of active galaxies \citep[e.g.,][]{2021ApJ...923...30L,2021ApJ...923...67H} with small viewing angles of their jets.
That is why, for the sake of simplicity, we call this a sample of blazars.

\section{Search for correlations with VLBI-bright
blazars}
\label{analysis}
In this section we perform statistical analysis of  neutrino data sets and VLBI-bright blazars. A neutrino data set can be the real observed data set (see  \autoref{neu_data}) or a Monte-Carlo simulated one.  We describe two types of analysis in \autoref{algorithms}. In each of them, we assess the significance of blazar-neutrino association,  either  using the blazars’ flux densities in the first type of analysis, or  using the number of coincident “blazar–neutrino” pairs in the second analysis.  We also present details of the Monte-Carlo algorithms utilized in these analyses. In \autoref{blind} we test which potential subset of cascade events can give the most significant result in the analyses and find that the high-energy neutrinos do. We also estimate the number of future events needed to reach a certain significance. We obtain p-values for the current neutrino data from Baikal-GVD and discuss the implications in \autoref{unblinding}.

\subsection{Monte Carlo algorithms}
\label{algorithms}
In order to analyze the possibility of finding directional correlation between cascade events and blazars and calculate test statistic, we create pseudo-experiment datasets with artificial randomized neutrino candidate events. Specifically, we assume a uniform distribution for the time of arrival events, and then implement a reshuffling of the times of all existing cascades for the given zenith and azimuth angles. Following this, we determine the right ascension and declination of the simulated events at the specified geographical coordinates of Baikal-GVD. The algorithm, which has been used to create such simulated datasets, is presented in Algorithm \autoref{alg:neutrino}.

\begin{algorithm}
\caption{Pseudo-experiment dataset generation}\label{alg:neutrino}

\begin{algorithmic}[1]
\State We take the horizontal coordinates, azimuth and zenith angle, of all 6157 cascade events from the data set, as well as their arrival times.
\State We reshuffle times for all events. To a good approximation, the sensitivity of the experiment only depends on azimuth and zenith angle.
\State Using the Baikal-GVD telescope coordinates ($51^\circ50^\prime$ N, $104^\circ20^\prime$ E), the original events coordinates and the new reshuffled times of arrival, we find the declination and right ascension of simulated events.
\State From the latter, we select a subsample for specific analysis. This may include, for example, the events below the horizon or high-energy neutrinos.
\end{algorithmic}
\end{algorithm}

To calculate the significance of coincidence of Baikal-GVD cascades with blazars, we perform analyses of two types (see \autoref{unblinding}). For this, two different statistics are introduced. Systematic errors of arrival directions are included in the error contours.

In the first type of analysis, we assess the significance using the blazars' flux densities following the procedure from \citet{Plavin2023}. 
Similarly, for blazars that fall inside the 90\% error regions of the neutrino candidate events, we average their flux density and take this value as a statistic. We take the geometric average, since the flux densities cover several orders of magnitude. Subsequently, we look if this value is higher than for randomly selected blazars. The second type of analysis includes finding the significance by the number of coincident ``blazar--neutrino'' pairs \citep[e.g.][]{2022icrc.confE1164A}. In this case, the number is taken as our statistic of interest. For the second type of analysis, we utilize a more conservative 50\% error region, as the 90\% error regions tend to have higher error values.

To calculate the significance, we use Monte-Carlo methods, which are implemented in the following Algorithm \autoref{alg:significance}.

\begin{algorithm}
\caption{Calculation of significance}\label{alg:significance}

\begin{algorithmic}[1]
\State We calculate a statistic of interest (either the average flux density or the number of coincident sources) using the coordinates of real cascade events and denote this statistic as $v_\mathrm{real}$.
\State We repeat $N=10000$ times the following:
\begin{enumerate}[label=\arabic*:,leftmargin=0.5cm]
  \item Create simulated cascade events as described in Algorithm~1.
  \item Calculate statistic $v_i$ for the simulated neutrinos.
\end{enumerate}
\State Count the number $M$ of simulated cascade events for which $v_i \geq v_\mathrm{real}$. The resulting $p$-value is then $p = \frac{M+1}{N+1}$.
\end{algorithmic}
\end{algorithm}

Given a certain assumption for the blazar neutrino signal, one can estimate the expected significance of the association between blazars and neutrinos for a given experimental data set. For this, we utilize pseudo-experiments generated from the time-scrambled Baikal-GVD data and determine what significance level can be potentially achieved in Algorithm \autoref{alg:significance} for specific sub-samples of cascade events. This is achieved as follows. First we generate fictitious sources of neutrinos (``blazar sky'') at positions of some of the pseudo-experiment's events. In this, real blazars from the RFC catalog are virtually ``shifted'' to the respective positions of the simulated events. The rest of the
pseudo-experiment's event sample is re-simulated many times, allowing us to calibrate the test statistic behavior. The details of the algorithm are presented in Algorithm \autoref{alg:potential_significance}.

\begin{algorithm}
\caption{Potential significance calculation}\label{alg:potential_significance}

\begin{algorithmic}[1]
\State We take (as in Algorithm \autoref{alg:significance}) one pseudo-experiment, which is created with the Algorithm \autoref{alg:neutrino} described before.
\State We repeat $N = 100$ times the following:
\begin{enumerate}[label=\arabic*:,leftmargin=0.5cm]
  \item Generate a random number $n$ from the binomial distribution $B(N_0,p_0)$ where $N_0$ is the total number of blazars and $p_0N_0$ is the expected rate of occurrences equal to the estimated number of blazar-associated neutrinos (see \autoref{blind} on how this number is determined).
  \item Select the given number $n$ of blazars from the part of the RFC catalog with a certain flux cut, which do not coincide with the pseudo-experiment neutrino cascades generated in the beginning.
  \item Change their coordinates so that they coincide with simulated cascade events within the error circles. For simplicity, we assume that each neutrino comes from a separate blazar.
  \item Count the total number of blazars, $n$, that coincide with cascade events within the error limits, and take this value as the $v_j$ statistic ($n$ shifted plus those from catalog that coincide with simulated cascades).
  \item Repeat $M = 1000$ times the following:
  \begin{enumerate}[label=\arabic*:,leftmargin=1cm]
    \item Create new realization of the background-only pseudo-experiment cascade events (using Algorithm \autoref{alg:neutrino}). The coordinates of blazars used in the outer loop are fixed in the inner loop.
    \item Calculate statistics $v_i$ for this new realization, where index $i$ refers to the inner loop.
  \end{enumerate}
  \item Count the number of cases $m_j$ for which $v_i \geq v_j$. Index $j$ refers to the outer loop. Then $p$-value for the signal-injected pseudo-experiment $j$ will be $p_j = \frac{m_j+1}{M+1}$.
\end{enumerate}
\State The average value $p = \frac{1}{N}\sum_{j}p_j$ is taken as a measure of the algorithm's sensitivity.
\end{algorithmic}
\end{algorithm}

The resulting expected $p$-value and corresponding significance depend on the simulated cascade events created at the beginning of Algorithm \autoref{alg:potential_significance} and can take on a wide range of values. To take this into account, we repeat the procedure described in this algorithm $K = 100$ times. Then we take the average number and report it as our result. 

\subsection{Monte-Carlo simulations and the expected signal}
\label{blind}

We perform Monte Carlo simulations for number of coincident blazars with Algorithm \autoref{alg:potential_significance}, described in the previous section. In this analysis, we take 1273 blazars with the 8 GHz flux density above 0.33 Jy (see \autoref{fig:HE_neu}). This cut-off for flux density turned out to be optimal and yielded the minimum $p$-value in the analysis of \citet{2021ApJ...908..157P}. It was concluded that the bright blazars above 0.33~Jy dominated the association with neutrinos from IceCube.  

We perform our optimization by testing which subset of neutrino candidate events can give us the most significant results. Specifically, we apply different cuts to cascade events based on their position in the sky, energy or hit multiplicity. In particular, we use high energy neutrinos with $E \geq 100$~TeV~(\autoref{tab:HE_events}), all neutrino candidate events and neutrino candidate events below the horizon level~(\autoref{tab:hor_events}). The 100~TeV threshold is commonly used in neutrino astrophysics \citep[e.g.][]{2014PhRvL.113j1101A,2018A&A...620A.174K,2019BAAS...51g.288G}. Such a threshold ensures a low level of background from atmospheric muons in all directions at once (see \autoref{tab:atm_ast_fl}) and it allows comparisons with previous analyses.

To estimate the number of astrophysical neutrinos that could come from blazars, we take into account the atmospheric background. In \autoref{tab:atm_ast_fl} we show the expected contributions of astrophysical and atmospheric neutrino fluxes for different subsets of the data \citet{Baikal-diffuse}. The difference between sum of astrophysical and atmospheric neutrino fluxes and total flux is due to atmospheric muons.
The astrophysical neutrino flux is normalised to value measured by IceCube \citep{2015PhRvD..91b2001A}.
We then use these numbers to find the number of blazars associated with neutrinos by multiplying the corresponding fraction of astrophysical signal and the total number of cascade events for a certain subset. To be more specific, we consider an extreme case in which we assume all astrophysical neutrinos originate from blazars, despite the potential existence of other neutrino-producing astrophysical sources. This
provides an optimistic limit to our results. 

\begin{table*}
\centering
\caption{Expected fractions of astrophysical and atmospheric neutrinos. Remaining fraction is due to atmospheric muons.}
\begin{tabular}{|r|c|c|c|c|}
\hline
  & $E > 10$ TeV & $E > 30$ TeV & $E > 60$ TeV & $E > 100$ TeV \\
\hline
  $N_\mathrm{hit} > 19$ astrophysical $\nu$: & 0.075 & 0.074 & 0.137 & 0.500 \\
  atmospheric $\nu$: & 0.104 & 0.054 & 0.053 & 0.116 \\
\hline
  $N_\mathrm{hit} > 15$ astrophysical $\nu$: & 0.020 & 0.042 & 0.089 & 0.257 \\
  atmospheric $\nu$: & 0.034 & 0.035 & 0.036 & 0.071 \\
\hline
  $N_\mathrm{hit} > 12$ astrophysical $\nu$: & 0.003 & 0.013 & 0.024 & 0.036 \\
  atmospheric $\nu$: & 0.006 & 0.014 & 0.014 & 0.014 \\
\hline
\end{tabular}
\label{tab:atm_ast_fl}
\end{table*}


It is assumed that the number of astrophysical neutrinos and atmospheric neutrinos under the horizon is equal to half of the total number of the corresponding neutrinos for energies $E > 10$ TeV and $N_\mathrm{hit} > 12$. This assumption was obtained using the zenith-angle distributions for atmospheric neutrinos and astrophysical neutrinos. Asymmetry of upward and downward events $A=\frac{N_{\rm up}-N_{\rm down}}{N_{\rm up}+N_{\rm down}}$ turned out to be 0.06 for atmospheric neutrinos and 0.02 for astrophysical ones.

\begin{figure}
\centering
\includegraphics[width=0.5\textwidth,trim=10mm 15mm 10mm 12mm]{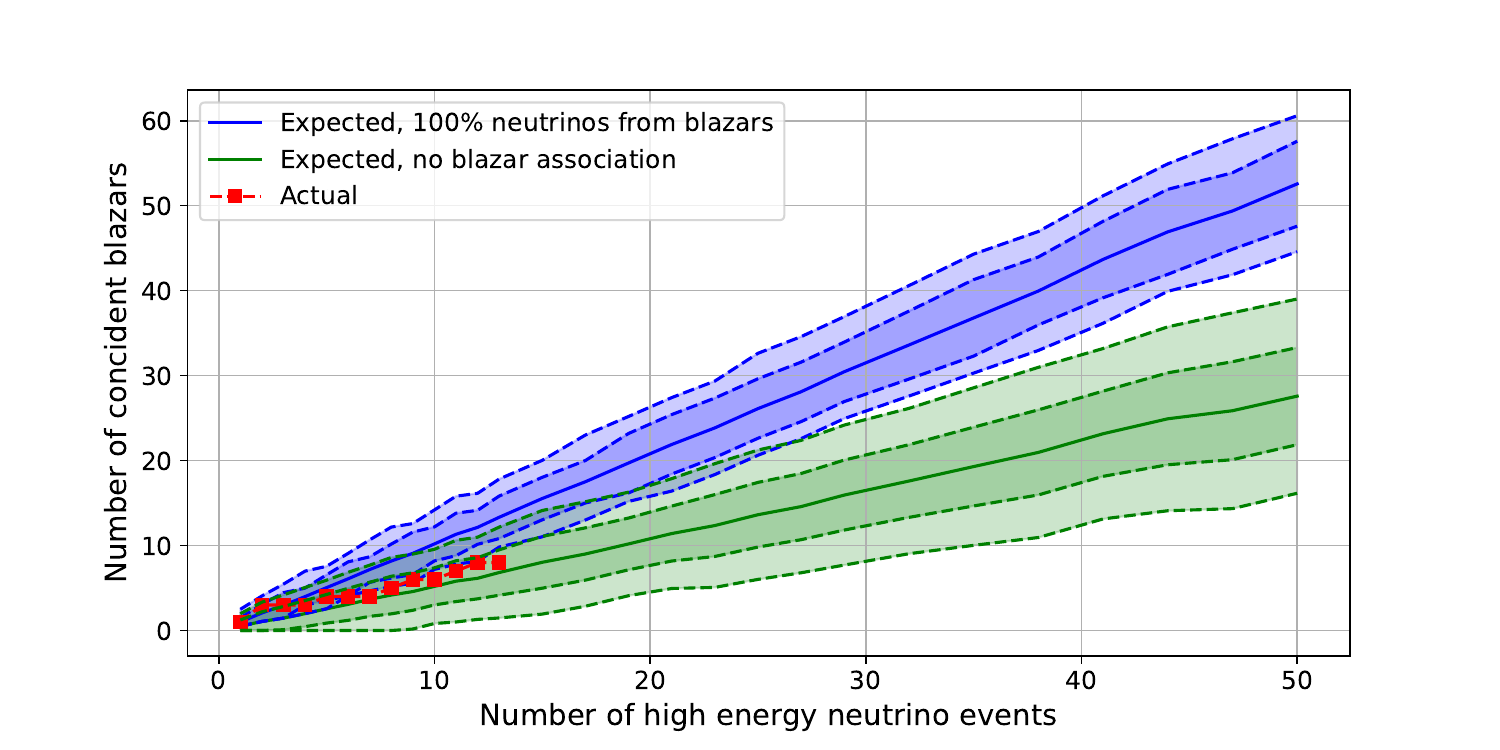}
\caption{Expected and observed numbers of blazars coincident with GVD cascades above 100~TeV versus the total number of these events in the sample. The blue line, and its 68\% and 95\% confidence intervals, depicted as blue shadowed regions, presents the expected number of pairs, obtained with Algorithm \autoref{alg:potential_significance} under assumption that blazars are responsible for 100\% of the astrophysical neutrino flux. In red, the actual number of coincident sources for real high-energy cascade events from \autoref{tab:HE_events} are shown in chronological order. The green line represents the expected number of pairs assuming there is no association between blasars and neutrinos. The corresponding 68\% and 95\% confidence intervals are shown as green shadowed regions. } 
\label{fig:sig_num}
\end{figure}

The summary of our results is given in \autoref{fig:sig_num} and \autoref{tab:signif}. Our simulations demonstrate that the best potential significance level can be achieved by taking the high-energy cascade events ($E > 100$~TeV). Such neutrinos provide the best significance since the expected fraction of astrophysical signal compared to the total one for them is the strongest. The reported 2.5$\sigma$ is the average significance of the
blazar-neutrino association expected for an experimental sample of 13 high energy events under the extreme assumption that the VLBI blazar sample is responsible for all of the diffuse astrophysical flux. If we find the standard deviation between different artificial neutrino realizations in Algorithm \autoref{alg:potential_significance}, the final significance and its uncertainty is 2.5$\sigma$ $\pm$ 1$\sigma$. 

\begin{table}
\centering
\caption{Expected significance levels of neutrino-blazar correlation obtained with pair counting, for the data sets used in this paper, assuming that all of the astrophysical flux comes from the VLBI blazars.}
\begin{tabular}{lrc}
 \hline
 Event sample  & Neutrino number & Expected significance \\
 \hline
 E$>$100 TeV & 13 &   2.5$\sigma$ $\pm$ 1$\sigma$  \\
 All events &  6157  &  1$\sigma$ \\
 Under horizon & 11 & 1.1$\sigma$ \\
\hline
\end{tabular}
\label{tab:signif}
\end{table}

We have also estimated how many events are needed to reach a certain significance. In \autoref{fig:sig_num} we present the number of coincident blazars with high-energy neutrino cascades and its 68\% and 95\% confidence intervals as a function of the number of events, assuming the 100\% dominance of blazars in the astrophysical neutrino flux. As one can see, in order to reach a 4$\sigma$ significance with high energy cascades, the experiment needs to accumulate about 50 events, a factor 4 more than is available to us now. Such number of events is expected to be detected by Baikal-GVD detector in around 4 years. This estimate does not take into account a constantly growing aperture of the detector.

\subsection{Actual data examination results}
\label{unblinding}

\subsubsection{Average flux density}

In this section, for the real Baikal-GVD neutrino cascades we perform the association-significance analysis using the average flux density of coincident blazars as the statistic. 
Such  analysis was introduced by \citet{2020ApJ...894..101P,Plavin2023} for track-like IceCube E > 200 TeV events.  
Similarly, we put a threshold  E > 200 TeV on Baikal-GVD cascades, which results in a very small sample of only 5 events. We employ the blazar catalog with 8-GHz flux densities exceeding 0.15 Jy, maintaining consistency with with the aforementioned paper. The resulting p-value turns out to be 0.18, consistent with expectations for this small sample (ten times smaller than the sample used by \citet{2020ApJ...894..101P,Plavin2023}.


In our study, we choose not to perform flux density optimisation on artificial data. The primary reason behind this decision is the inherent complexity of the flux model of shifted blazars. Estimating the appropriate flux values to assign to blazars in artificial data proves to be a challenging task due to the intricate nature of the model. As a result, we opt to use real data exclusively when working with average flux density significance analysis.

\subsubsection{Pair counting results}
The significance of association between blazars and high-energy neutrinos utilizing the actual data is found also using the number of neutrino-blazar coincidences as the statistic. For this,  neutrinos with energies above 100 TeV (13 events) are included. We use Algorithm \autoref{alg:significance} described in \autoref{algorithms} and take the number of ``blazar--neutrino'' pairs as the statistic of interest. This analysis yields no
significant detection. The corresponding $p$-value is about 0.35 (statistical significance less than 1$\sigma$). The actual number of neutrino-blazar coincidences is close to the average number of coincidences in a typical background-only pseudo-experiment. Although our result falls below the estimated 2.5$\sigma$ level as mentioned in \autoref{blind}, this significance might still arise by chance (see \autoref{fig:sig_num} and its uncertainty region). Therefore, our findings do not necessarily conflict with each other. Simultaneously, the source of this discrepancy might be linked to an overestimation of the astrophysical neutrino flux in our data associated with blazars, potentially due to the inclusion of flux from other astronomical sources. Future studies will aid in discerning between these two possible explanations.

\begin{figure}
\centering
\includegraphics[width=0.5\textwidth, trim=0cm 1.2cm 0cm 0cm]{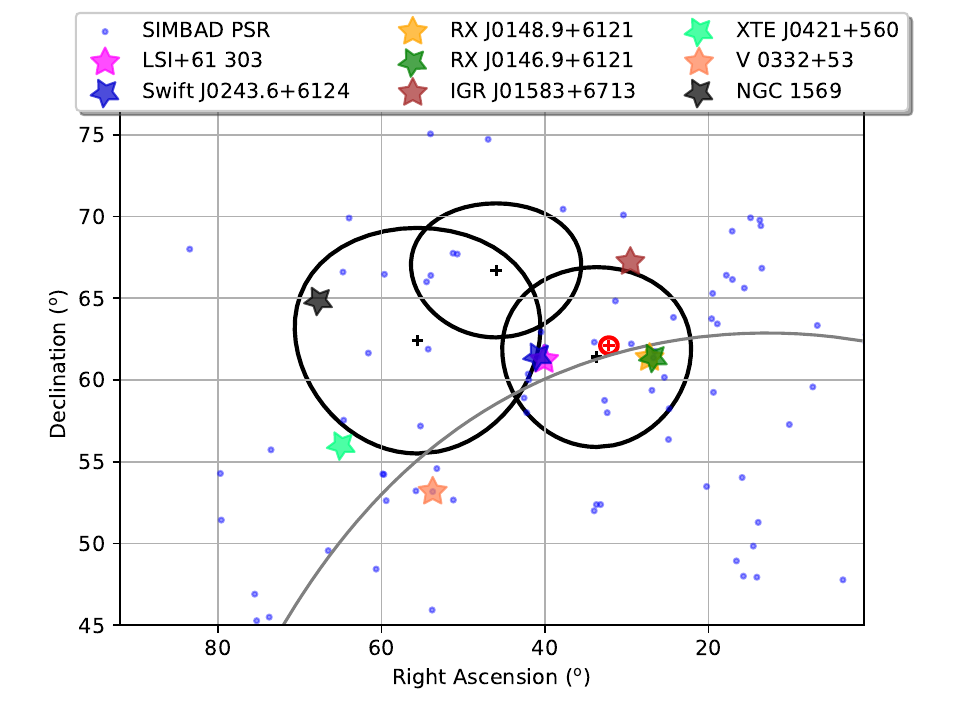}
\caption{Three high energy cascade events (GVD190216CA, GVD190604CA and GVD210716CA from \autoref{tab:HE_events}) close to the Galactic plane (grey line) and their corresponding 90\% errors (black). The  persistent northen-hemisphere maximum of the IC probability map over  2008--2015~\citep{2017ApJ...835..151A} is shown by the red plus and circle, drawn as the uncertainty $0.5^\circ$ at 90\% level.   
Pulsars from the SIMBAD Astronomical Database are shown as dots. Several close high-mass X-ray binaries are indicated.} 
\label{fig:triplet}
\end{figure}

\section{Promising coincident sources}
\label{special}

In this section, we scrutinize potential coincident candidates associated with the Baikal-GVD cascades, incorporating both Galactic and extragalactic ones. A brief summary is presented in \autoref{tab:summary}.

\begin{table*}
\centering
\caption{Summary of the promising neutrino-coincident Galactic and extragalactic sources. See \autoref{galactic} and \autoref{extragalactic} for more details.}
\begin{tabular}{llll}
\hline
Source & Type & Key Features & Baikal Cascade Events \\
\hline
\lsi{} & Binary system & Flares in wide electromagnetic spectrum & GVD190216CA, GVD190604CA \\
\pulx{} & X-ray Pulsar & Ultraluminous X-ray source, super-Eddington luminosity & GVD190216CA, GVD190604CA \\
\hline
2023$+$335 & Blazar & 2~Jy VLBI flux density, gamma-ray detection by \textit{Fermi} LAT & GVD210409CA \\
2021$+$317 & Blazar & 1.5~Jy VLBI flux density, flat spectrum & GVD210409CA \\
2050$+$364 & Blazar & 1.2~Jy VLBI flux density, steep spectrum & GVD210409CA \\
0529$+$075 & Blazar & 1.3~Jy VLBI flux density, variable spectrum & GVD210418CA \\
0506$+$056 & Blazar &  Temporally coincident radio flare & GVD210418CA \\
0258$-$184 & Blazar & Temporally coincident radio flare & GVD190523CA, GVD210501CA \\
1935$-$179 & Blazar & Temporally coincident radio flare & GVD200826CA \\
\hline
\end{tabular}
\label{tab:summary}
\end{table*}

\subsection{Galactic sources}
\label{galactic}

Already first IceCube data had demonstrated possible flux from the Galactic ridge consistent with \textit{Fermi} LAT flux at 100 GeV energies \citep{Neronov:2013lza}. However, the arrival directions of the early IceCube events were subsequently reconsidered. Assuming a certain model of cosmic-ray interactions with interstellar gas, the global contribution of Galactic disk to IceCube and ANTARES data was found not to exceed $\sim 10\%$ of the total astrophysical neutrino flux \citep{ANTARES:2018nyb}. Recently, in a model-independent search for the Galactic anisotropy, a significant excess of IceCube track-like events from low Galactic latitudes was found, consisting with the Galactic contribution of about 30\% of the total flux above 200~TeV \citep{neutgalaxy}. 
Subsequently, ANTARES reported a hint of the neutrino emission associated with the Galactic ridge \citep{ANTARES-ridge}. These results were later supported by an IceCube analysis of cascade data \citep{IC_Galaxy2023}.


Using the Baikal-GVD high-energy cascade sample, it was noted that the error circles (90\% CL) of arrival directions of three events, see \autoref{tab:HE_events}, intersect, forming a triplet close to the Galactic plane \citep[][]{Baikal-diffuse}. We estimate the $p$-value of finding such a triplet on the sky using Monte Carlo simulations. Specifically, we use Algorithm \autoref{alg:neutrino} and slightly modified version of Algorithm \autoref{alg:significance}, for which our statistics of interest is equal to the number of triplets present in the simulated events. The final $p$-value turns out to be equal to 0.024 (2.26$\sigma$). While this result is not highly significant (falling short of the 3$\sigma$ standard often used in the scientific community), it nonetheless presents an intriguing implication. This intersection and the computed p-value of 0.024 can be viewed as a suggestive indicator, pointing towards this specific region near the Galactic plane as being of potential interest. Such hints invite future investigations to delve deeper into this area. We add that the intersection of two of them contains very well-known Galactic sources of high-energy radiation, \lsi{} and \pulx, see \autoref{fig:triplet}.

\lsi is a binary system of compact object and companion star located at $2.6\pm0.3$~kpc from the Solar system ~\citep{gaia2018Cat}. The J2000.0 coordinates for this source are  40\dotdegrees131917, 61\dotdegrees229333 and its location is consistent with two cascade events (events GVD190216CA and GVD190604CA in \autoref{tab:HE_events}). 
It is known to be variable, with flares coming periodically in a wide range of the electromagnetic spectrum from radio to TeV.  
The compact object in this system is most probably a  pulsar, which is strongly indicated by pulsations with a period of $P=269$~ms ($>20\, \sigma$) observed recently  by FAST \citep{Weng+2022}. 

The orbital period of the binary system of \lsi is $26.496$~d. 
Additionally, flare brightness modulations are observed in all energy ranges with a longer period $\sim 1659$~d~\citep{Jaron+2021}. 
The Baikal-GVD events came at phases 0.3037 and 0.4098
of the orbital period and 0.1403 and 0.2059
of the major period 
(with zero phase at 43366.275 MJD). 
\cite{Prokhorov-Moraghan22}  report that the latest of the high $\gamma$-ray states of \lsi, most
likely associated with superorbital modulation, started
in March 2019 and lasted for 96 weeks.

Almost in the same position of sky there is another interesting object, the Galactic X-ray pulsar \pulx~\citep[$2-6$~kpc,][and references therein]{2021AJ....161..147B,Reig+2020}.
Its coordinates are  40\dotdegrees918437, 61\dotdegrees434377. 
The source was discovered in 2017 when a giant flare was taking place in it. 
Currently, this is the only discovered pulsating ultraluminous X-ray source (PULX) in the Galaxy.

The super-Eddington luminosity of PULX sources ($\sim 10^{39}-10^{41}$~erg/s) is explained by virtue of the intense accretion of matter from a neighboring massive star onto a magnetized neutron star
~\citep{Mushtukov-Tsygankov22,King+2023}.
Although radio emission from such sources is generally not observed, it should be noted that a radio jet~\citep{Eijnden+2018} was observed from \pulx
at the time of the X-ray peak in 2017. 
The spin period of the neutron star is 9.86~s~\citep{2017ATel10812....1J}. 
The period of the binary is 28.3 d~\citep{2018A&A...613A..19D}. 
In 2019, the source underwent another X-ray outburst, about 6 times dimmer than that of 2017. GVD190216CA happened at its end, and according to \citet{Liu+2022} who have analyzed the long-term evolution of optical fluxes in $V$, $I$, and $R$ band, the mid of February 2019 is the time of maximal dissipation of the Be-star's decretion disc.

We have analysed available archival data in X-ray and gamma from \lsi and \pulx at times of the two Baikal-GVD events. 
The Swift/BAT 15-50 keV daily average light curve shows a flux increase from \lsi a few days before the neutrino GVD190604CA. 
Formally, this was the largest flux from \lsi for the entire observation period since 2005.  
However, at the time Swift observed the sky area without `dithering' mode,  which indicates a low significance of data.
Moreover, around the 140th day of 2019, Swift/BAT shows some increase in daily counts for \pulx as well, probably  indicating the fluctuation not intrinsic to the source.
MAXI, which does not resolve \lsi and \pulx being separated by 25$^\prime$45.93$^{\prime\prime}$, also shows a moderate (twice the background) 2-20~keV flux enhancement in June of 2019.
\textit{Fermi} LAT 0.1-300~GeV data indicates enhanced (about two-fold) radiation at the beginning of June 2019 compared with the long-term period-stacked light curves.
At the time of the third event from the triplet, GVD210716C, Swift/BAT, MAXI, or HMXT data show no suspicious features for \lsi and \pulx. 
Thus, existing data does not allow a certain conclusion about the association of any triplet cascade events with high-energy electromagnetic radiation from either of the sources.

Recall that only about 50\% of events in \autoref{tab:HE_events} are expected to come from astrophysical objects, while another half are due to background. Thus, two events, GVD190216CA and GVD190604CA,  coming from direction to same astrophysical source, \lsi  or \pulx (or both),
would make this direction the brightest neutrino source in the sky providing
a large fraction 
of the total astrophysical neutrino flux. 

Though a neutrino source at this Northern declination should be hardly accessible by IceCube at highest PeV energies because of the Earth opacity,
still it would be visible at energies $\sim 100$~TeV and lower.
Intriguingly, the position of one of events in  the Baikal-GVD triplet in the sky coincides with the highest-significance Northern hot spot in the IceCube 7-year sky map \citep{2017ApJ...835..151A} (see \autoref{fig:triplet}), though other directions bring higher significances in subsequent similar analyses \citep{IceCube:10yr,doi:10.1126/science.abg3395}.
The corresponding source has not been reported by IceCube at $E>100$~TeV, so its observed  
flux in Baikal GVD may be an upward fluctuation, and the real flux of this source should be smaller. Notice that, while a continued neutrino emission from the source is not observed by IceCube, a flaring source cannot be ruled out at the moment from a theoretical standpoint~\citep[e.g.][]{2009PhRvD..79d3013N,Bykov:2021maf}. 
%
At the same time, the arrival of three unrelated high-energy neutrinos from directions close the Galactic plane is also possible and consistent with the recent finding that the IceCube high-energy neutrino flux has a Galactic component \citep{neutgalaxy,IC_Galaxy2023}.

\subsection{Extragalactic sources}
\label{extragalactic}

Following \cite{2020ApJ...894..101P,Plavin2023}, we select blazars with the brightest radio emission as potential associations with high-energy cascade events. See \autoref{data_vlbi} for details on the radio source sample collected on the basis of VLBI observations.

There are four blazars with historical VLBI flux density at 8~GHz above 1~Jy that coincide with the Baikal-GVD high-energy cascade events (\autoref{tab:HE_events}). Three of them are close to event GVD210409CA. The brightest is 2023$+$335 ($z = 0.22$): its average VLBI flux density is 2~Jy at 8~GHz, with a flat or slightly inverted spectrum at GHz frequencies. Moreover, 2023$+$335 is detected by \textit{Fermi} LAT in gamma rays \citep{2012ApJ...746..159K}. Two others are 2021$+$317 ($z = 0.36$) with 1.5~Jy VLBI flux density at 8~GHz and a flat spectrum as well as 2050$+$364 ($z = 0.35$) with 1.2~Jy, a steep spectrum, and a lower dominance of parsec-scale structures. The fourth bright blazar is 0529$+$075 ($z = 1.25$, \citealt{2005ApJ...626...95S}) coincident with the event GVD210418CA. Its VLBI flux density is 1.3~Jy at 8~GHz, it has a highly variable spectrum, ranging from falling to inverted.

Additionally, there are several coincidences that show evidence of temporal correlation: an ongoing major radio flare when neutrino is detected from the direction of the source (see \cite{2022JETP..134..399A} for more details). Such behavior is characteristic to neutrino-associated blazars, see \cite{2020ApJ...894..101P,2021A&A...650A..83H,Plavin2023}. The most notable is the flaring blazar TXS~0506$+$056 ($z=0.34$, average 0.5~Jy at 8~GHz) coincident with the event GVD210418CA from \autoref{tab:HE_events}. This association is investigated further in a dedicated paper \citep[][]{2022arXiv221001650B}. 
Other blazars with hints for temporally coincident flares are 0258$-$184 (events GVD190523CA and GVD210501CA), and 1935$-$179 (event GVD200826CA). These objects were highlighted and discussed in \cite{2022JETP..134..399A} together with their radio light curves.

Apart from that, some of the sources discussed above can be further examined in the results of VLBA monitoring observations, including kinematics measurements and VLBI-\textit{Fermi} analysis in, e.g., \citet{2021ApJ...923...30L,2022MNRAS.510..469K}.
More dedicated single-dish and VLBI monitoring observations are needed to achieve the required level of significance of the neutrino-blazar analysis \citep[see discussions in ][]{2022A&A...666A..36L,YYK_ngEHT23}.




\section{Summary}
\label{sum}
In this paper, we discuss potential associations of Baikal-GVD cascade events with astrophysical sources. Compared to ice, the use of liquid water allows for better angular resolution, thus making neutrino astronomy with cascades possible. In particular,
we have analyzed directional association between VLBI-selected radio blazars and Baikal-GVD events. Monte-Carlo simulations on artificial datasets demonstrate that, among the standard Baikal-GVD cascade data sets, the best significance levels for this association could be achieved by using events with $E>100$~TeV, if the number of coincident ``blazar-neutrino'' pairs is used as the test statistics.
We estimate the expected signal as a function of the dataset size. The first set  of the April 2018 -- March 2022 events demonstrates no significant effect, in agreement with our simulations. However, with the present low statistics, we cannot exclude also the overestimation of the number of neutrinos associated with blazars. This overestimation could arise due to other astrophysical sources of neutrinos. Future data could help in resolving this problem.

Besides pair counting, we have also performed an analysis of the statistics based on the average flux density of coincident sources and events with $E\ge200$~TeV. Such analysis represents the most sensitive test  \citep{2020ApJ...894..101P,Plavin2023}. There are only 5 such events in the currently available Baikal-GVD sample. As expected for a small sample, no statistically significant correlations are found, $p$-value being equal to 0.18. 

We have listed and discussed promising coincident high-energy astrophysical sources, including Galactic and extragalactic ones. 
The most notable Galactic sources turned out to be \lsi and \pulx, which fall within the 90\% uncertainty regions of two events. 
The extragalactic sources that caught our attention are 2023$+$335, 2021$+$317, 2050$+$364, and 0529$+$075, which have high values of their average flux densities. We note that TXS~0506$+$056, 0258$-$184 and 1935$-$179 demonstrated temporal coincidences of radio flares and times of neutrino arrival. 

The overall results indicate that increasing the number of cascades and use of track-like events are necessary for firm astrophysical conclusions. This will be achieved in the nearest future with continuously growing aperture and improving sensitivity and reconstruction methods of the Baikal-GVD neutrino telescope \citep[e.g.][]{2021PAN....84.1600A,2022JETP..134..399A}.

\section*{Acknowledgements}

We thank Eduardo Ros for useful comments on the manuscript. We thank Hans Krimm for advice concerning Swift data.
%
This work is partially supported by the European Regional Development Fund-Project ``Engineering applications of microworld physics'' (CZ 02.1.01/0.0/0.0/16 019/0000766).

\section*{Data Availability}

 The article uses neutrino data. This paper presents the parameters for the high-energy cascades, as well as those for the under-horizon cascades, in conjunction with \citet{Baikal-diffuse}. The data for all other cascades underlying this article will be shared on reasonable request to the corresponding author.
The analysis utilizes the publicly available Astrogeo\footnote{\url{http://astrogeo.org/vlbi_images/}} database and the Radio Fundamental Catalogue\footnote{\url{http://astrogeo.org/sol/rfc/rfc_2023a/}}.



\bibliographystyle{mnras}
\bibliography{baikal_neutrino} 






\bsp	
\label{lastpage}
\end{document}